\documentstyle[epsf,psfig]{aipproc}

\begin{document}

\title{Software Corrections of Vocal Disorders}

\author{Lorenzo Matassini}

\address{Max-Planck-Institut f\"ur Physik komplexer Systeme\\
         N\"othnitzer Str.\ 38, D 01187 Dresden, Germany\\
         email: lorenzo@mpipks-dresden.mpg.de}

\author{Claudia Manfredi}

\address{Dept. of Electronic Engineering - Univ. of Florence\\
         via Santa Marta 3, I 50139 Firenze, Italy\\
	 e-mail: manfredi@die.unifi.it}

\maketitle

\begin{abstract}
We discuss how vocal disorders can be post-corrected via a simple nonlinear noise reduction scheme. 
This work is motivated by the need of a better understanding of voice dysfunctions. This would entail a twofold 
advantage for affected patients: Physicians can perform better surgical interventions and on the other hand 
researchers can try to build up devices that can help to improve voice quality, i.e. in a phone 
conversation, avoiding any surgigal treatment.
As a first step, a proper signal classification is performed, through the idea of geometric signal separation 
in a feature space. Then through the analysis of the different regions populated by the samples coming from healthy 
people and from patients affected by T1A glottis cancer, one is able to understand which kind of interventions 
are necessary in order to correct the illness, i.e. to move the corresponding feature vector from the sick region 
to the healthy one.
We discuss such a filter and show its performance. 

\vspace*{4mm}

\vspace*{4mm}

\end{abstract}

\section*{Phase Space Reconstruction}

\begin{figure}
\centerline{\psfig{file=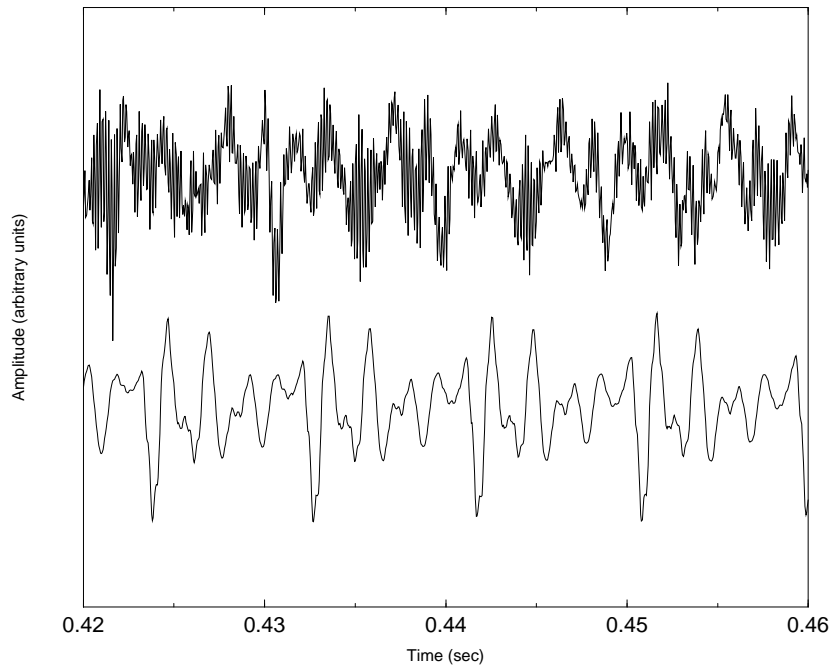,width=11cm,angle=270}}
\caption[]{\small\label{fig:esempio01}
Typical shapes of the investigated time series. Upper panel: Dysphonic voice. Lower panel: Healthy sample.}  
\end{figure}

The first simple difference between dysphonic and healthy voices is shown in Fig.\ref{fig:esempio01}, where the time 
evolution of the amplitude of a microphone-registered sound is represented. The upper panel could be interpreted as a 
highly noisy time series, but careful investigations reveal that this is not the case: Applying a simple low-pass filter 
would only introduce a distortion bigger than the original noise level. Some of the noise-like structures belong to 
the time series and one has to be able to correctly identify what is worth keeping and what has to be eliminated during 
the correction procedure.

From a theoretical point of view, this paper relies on the theory of dynamical systems and deterministic chaos; the former
implies that the time evolution is defined in some phase space, the latter offers a striking explanation for irregular 
behaviour and anomalies in systems which do not seem to be inherently stochastic. 
Even very simple chaotic dynamical systems can exhibit strongly irregular time evolution without random inputs.

Consider for a moment a purely deterministic system. Once its present state is fixed, the states at all future times
are determined as well. Thus it will be important to establish a vector space, the so-called \emph{phase space}, for
the system such that specifying a point in this space specifies the state of the system, and vice versa. This implies that
we can study the dynamics of the system by studying the dynamics of the corresponding phase space points.

Unfortunately what we observe in an experiment is not a phase space object but a time series, most likely only a
scalar sequence of measurements. We therefore have to convert the observations into state vectors: This is the
problem of \emph{phase space reconstruction} which is technically solved by the method of delays or related
constructions \cite{takens,sauer,kantz1}. 
Most commonly, the time series is a sequence of scalar measurements of some quantity which depends on
the current state of the system, considered at multiples of a fixed sampling time:
\begin{equation}
s_n=s({\bf x}(n\Delta t))+\eta_n,
\end{equation}
namely we look at the system through some measurement function $s()$ and make observations only up to some random
fluctuations $\eta_n$, the measurement noise. A \emph{delay reonstruction} in $m$ dimensions is then formed by the
vectors $\beta_n$, given as
\begin{equation} \label{eq:embedd}
\beta_n=(s_{n-(m-1)\nu},s_{n-(m-2)\nu},...,s_{n-\nu},s_n).
\end{equation}
The time difference in number of samples $\nu$ or in time units $\nu \Delta t$ between adjacent components of the
delay vectors is referred to as the \emph{lag} or \emph{delay time}.

A number of embedding theorems are concerned with the question under which circumstances and to what extent the
geometrical object formed by the vectors $\beta_n$ is equivalent to the original trajectory ${\bf x}_n$. Here
equivalent means that they can be mapped onto each other by a uniquely invertible smooth map and, under quite general
circumstances the attractor formed by $\beta_n$ is equivalent to the attractor in the unknown space in which the
original system is living if the dimension $m$ of the delay coordinate space is sufficiently large.

\section*{Recurrence plots of the voice}

Human voices form an aperiodic and highly nonstationary signal. A sentence can be decomposed in subunits, called
phonemes, which can be considered as different types of dynamics. Careful investigation of time and length scales
shows that the sound wave characterizing a single phoneme (duration between 50 and 200 ms) has a characteristic
profile (\emph{pitch}) of about 5-15 ms. A kind of phase angle on this highly nontrivial oscillation will then 
identify the instantaneous amplitude. Thus a delay reconstruction should allow us to identify the actual phoneme 
and the phase inside it.

\begin{figure}
\centerline{\psfig{file=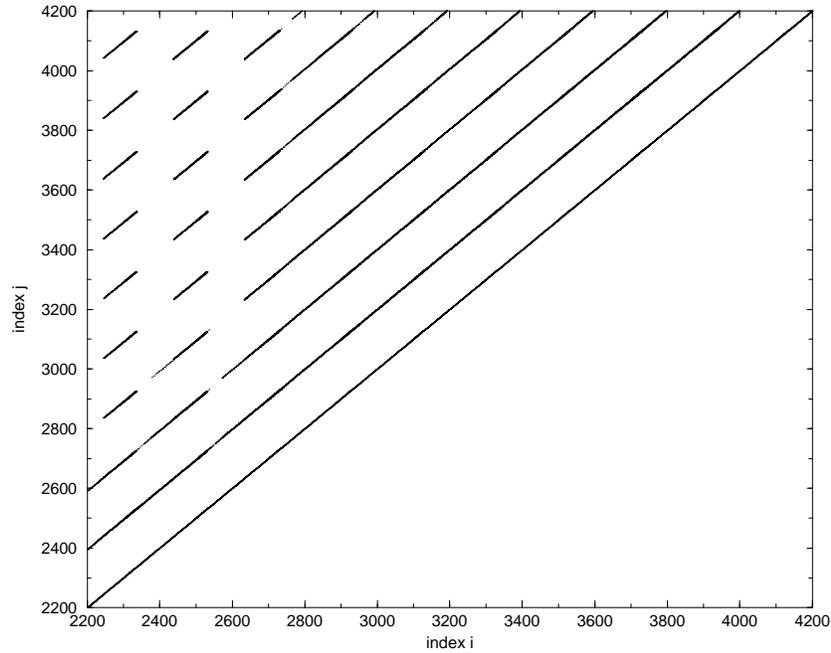,width=11cm,angle=270}}
\caption[]{\small\label{fig:recurrence}
Section of a recurrence plot of the time series in the lower panel of Fig.\ref{fig:esempio01}. A recurrence plot is 
constructed on the basis of mutual distances between points belonging to the same trajectory. Every dot in $(i,j)$
implies that $|\beta_i - \beta_j| < \epsilon$. There are almost no dots for $|i-j| > 2000$, reflecting the lack of 
interphoneme similarities (for this particular choice of $\epsilon$). The occurrence of lines mirrors the deterministic
character of the system: It is therefore possible to distinguish between different dynamical regimes, namely to
introduce pseudostates in the system.}  
\end{figure}

Looking at the Eq.\ref{eq:embedd} one realizes that at least two parameters are involved in the delay reconstruction
from a scalar time series, namely $m$ and $\nu$. Some recipes for an optimal tuning of them are available at the moment, 
but an adequate theory is still missing; in any case, a good tool is
represented by the \emph{recurrence plot} (see \cite{recurrenceplot,recurrenceplot2}). This method was used for the first
time to study recurrencies and nonstationary behaviour occurring in dynamical systems. It allows to identify system properties
that cannot be observed using other linear and nonlinear approaches and it is especially useful for analysis of nonstationary
systems with high dimensional and/or noisy dynamics.

Recurrence plots are constructed on the basis of mutual distances between point belonging to the same trajectory. In the 
plane of indices $i$ and $j$ a dot is printed whenever the delay vectors $\beta_i$ and $\beta_j$ fulfill the relation 
$|\beta_i - \beta_j| < \epsilon$. So a recurrence plot depends also  on the parameter $\epsilon$.
The Fig.\ref{fig:recurrence} proves that our delay vectors really represent meaningful states, where the
line structure shows the approximate periodicity inside the phonemes and the number of intraphoneme neighbours. 

The distance between two consecutive lines indicates the duration of the basic structure (\emph{pitch}) inside a phoneme. 
We can see from Fig.\ref{fig:recurrence} that this involves approximately 200 points. Considering that the sentence has been
sampled at a 22,05 kHz rate, this implies a profile of about 9 ms. In order to correctly identify such a structure,
in the delay reconstruction of Eq.\ref{eq:embedd}, the product $\nu m$ (the so-called \emph{time window}) has to be
bigger than 200. A more detailed insight into the role of this two parameters is given in Ref. \cite{dimitris}.

\section*{Nonlinear noise reduction}

Noise reduction means that one tries to decompose a time series into two components, one containing the signal, the 
other random fluctuations. Implicitly we always assume that the data represents an additive superposition of two different 
components which have to be distinguishable through some objective criterion.

The classical statistical tool for obtaining this distinction is the power spectrum. Random noise has a flat, or at
least a broad, spectrum, whereas periodic or quasi-periodic signals have sharp spectral lines. After both components
have been identified in the spectrum, a Wiener filter can be used to separate the time series accordingly. This
approach fails for our purposes here because the undesirable part of the signal is not what is usually considered to be
noise. It is very strongly correlated to the clean part of the signal, indeed it is part of the signal.
Even if parts of the spectrum can be clearly associated with the signal, a separation into signal and noise fails for
most parts of the frequency domain.

The filter we use has been proposed in \cite{lms} and arises from the chaotic deterministic systems field, where
the determinism yields a criterion to distinguish the signal and the noise (which is supposed not to be
deterministic). Let the time evolution of the signal be deterministic with an unknown map $f$. All that we have
knowledge of are noisy measurements of this signal:

\begin{equation}
s_n=x_n+\eta_n, \hspace*{1cm} x_n=f(x_{n-m},...,x_{n-1}).
\end{equation}
Here $s_n$ is the scalar time series we can measure, $x_n$ the clean signal and $\eta_n$ the superimposed noise.
In order to obtain an estimate $\hat{x}_{n_0-m/2}$ for the value of $x_{n_0-m/2}$ we form delay vectors
\begin{equation}
{\bf s}_n=(s_{n-m+1},...,s_n) 
\end{equation}
and determine those which are close to ${\bf s}_{n_0}$. The average value of $s_{n-m/2}$ is then used as a cleaned
value $\hat{x}_{n_0-m/2}$:

\begin{equation} \label{eq:filtro}
\hat{x}_{n_0-m/2}=\frac{1}{|U_{\epsilon}({\bf s}_{n_0})|} \sum_{{\bf s}_n \in U_{\epsilon}({\bf s}_{n_0})} 
                  s_{n-m/2}.
\end{equation}
Here $|U_{\epsilon}({\bf s}_{n_0})|$ denotes the number of elements of the neighbourhood 
$U_{\epsilon}({\bf s}_{n_0})$ of radius $\epsilon$ around the point ${\bf s}_{n_0}$, which is never empty, no matter
how small a value of $\epsilon$ we choose: It always contains at least ${\bf s}_{n_0}$. This is good to know since we
have to make some choice of $\epsilon$ when we use this algorithm. It is guaranteed that if we choose $\epsilon$ too
small the worst thing that can happen is that the only neighbour found is ${\bf s}_{n_0}$ itself. This, however,
yields the estimate $\hat{x}_{n_0-m/2}=s_{n_0-m/2}$ which just means that no correction is made at all.

To get an impression of how the local projective noise reduction scheme works, assume that one has to eliminate noise
from music stored on an old-fashioned long playing record, induced by scratches on the black disc. The task becomes
almost trivial if one can make use of several samples of this LP. When playing them synchronously, the signal part of
the different tracks id identical, whereas the noise part is independent: As a consequence of that, already a simple
averaging would enhance the sound quality. In deterministic chaotic signals, this redundancy is stored in the past:
Similar initial conditions will behave in a similar way, at least for short periods. In human voice signals there is
no need to suppose a chaotic behavior, since every phoneme is made up of pitches in an almost periodic fashion. This
means that every logical unit provides all the redundancy required for its filtering.

How can we make sure that all we do is to reduce noise without distorting the signal? In Fig.\ref{fig:esempio01} two 
typical time series, one related to a healthy voice, the other to a sick one, are compared. In the lower panel we can 
see the recurrence of structures of length approximately 9 ms inside a phoneme, in total agreement with the recurrence 
plot analysis. 
The person who has spoken this sentence is absolutely healthy and the corresponding signal looks very clean.
The same repetition is not that clear in the upper panel, where a patient affected by a T1A glottis cancer was asked
to say the same group of words in the same environmental conditions, namely in a professional quiet room. The signal
looks very noisy, but listening to it one perceives only the sick voice and no indication of noise, at least of what
one usually means saying noise.

Spectral analyses show the same result: The vocal disorder reveals itself as a noisy signal, but the nature of this
noise is not the conventional one. It is not addivite and it is correlated with the other sub-part of the sentence
one would like to isolate. In \cite{herzel2,ishizaka,manfredi,reuter,pinto,kumar} it is shown how some of the 
complexities observed in disordered voices
are not caused by random external input to the vocal apparatus, but by the intrinsic nonlinear dynamics of vocal
fold movement. Normal phonation corresponds to an essentially synchronized motion of all vibratory modes. A change of
parameters such as muscle tension or localised vocal fold lesions may lead to a desynchronization of certain modes
resulting in the appearance of these new features that look like noise.

The distinction of the signal in a deterministic part and in a stochastic component is then not possible.
Nevertheless one would still like to extract the main structure from the signal, suppressing the noise-like features.
For this purpose the previously described filter is worth applying to correct the disorders: If we choose
suitably the involved parameters we are able to identify the structures inside a phoneme and to perform an average of 
them via Eq.\ref{eq:filtro}. Every point belonging to a 
structure is replaced by a local average of similar points coming from different structures (this is done searching
for neighbours in the embedding space). The resulting signal does not show just an exact repetition of these parts
because the averaging procedure acts only locally and the points involved in the computation varies from structure
to structure. Also in a normal signal, in fact, the repetitions are not perfect. 

\begin{figure}
\centerline{\psfig{file=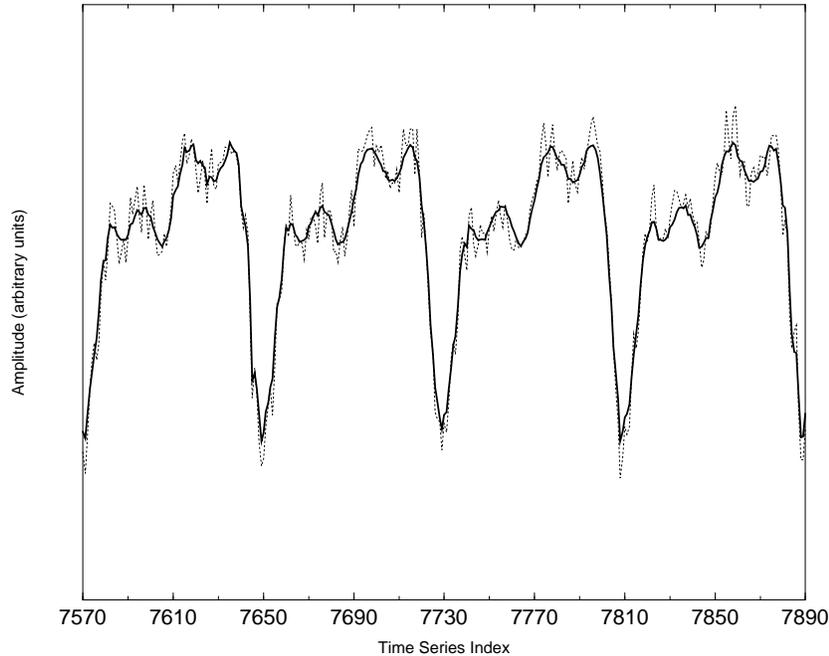,width=11cm,angle=270}}
\caption[]{\small\label{fig:esempio02}
Corrections performed by the nonlinear noise reduction to a dysphonic voice. Dotted line: Original voice. Solid line: 
Corrected voice.}  
\end{figure}

One has then to be careful when choosing the $\epsilon$: A too big value would result in a drastic averaging and the 
resulting signal would sound too artificial. On the other hand a small value of $\epsilon$ is not able to perform any 
correction. In Fig.\ref{fig:esempio02} we show what happens when using the filter with a \emph{good} set of parameters: 
The main shape of the sub-structures is preserved, but the noise-like features are attenuated. 
The meaning of good here is the following: The recurrence plot of Fig. \ref{fig:recurrence} shows that the system
generating the time series we are analyzing really possesses dynamical regimes. Since we want to be able to
distinguish between different states, the vectors $\beta_n$ of the delay reconstruction has to comprise all the
points belonging to a pitch. So the product $\nu m$ of the lag and the embedding dimension has to be approximately 200
in the example depicted on Fig. \ref{fig:recurrence}. A choice of a big $m$ results on big computational efforts; on
the other hand a big $\nu$ produces worse results. We refer again to Ref. \cite{dimitris} for more details.
The parameter $\epsilon$ is used for the neighbourood relations between vectors in the embedding space. We draw a point
in $(i,j)$ on the recurrence plot if $|\beta_i - \beta_j| < \epsilon$. The value of $\epsilon$ has to be bigger than the
noise level. A good way to tune it is through the recurrence plot: Whit a small $\epsilon$ we get almost no recurrences,
a very big $\epsilon$ is such that the relation $|\beta_i - \beta_j| < \epsilon$ is fulfilled from almost all the vector
pairs. The optimal value of $\epsilon$ is mirrored on a recurrence plot where the lines are as long as possible, as thin
as possible and all the recurrences belong to lines.

If we tune the parameters according to this recipe, the filtered signal sounds more \emph{normal} than the original,
even if some characteristic aspect of the voice has been lost. In the horizontal axis of Fig.\ref{fig:esempio02} we report 
the index of the time series in order to show the number of points involved in the filtering procedure. Every structure 
lasts 80 points; in a space of such a dimensionality, the vector having these elements as components is represented as a 
single point. The four structures of Fig.\ref{fig:esempio02} are neighbours because the involved sequences of points are 
approximately the same. As shown in \cite{dimitris}, it is not necessary to consider such a big space, but one can skip 
some intermediate point, taking care to cover the full structure in any case.

\section*{Results}

We are dealing with three categories of subjects: (i) patients affected by T1A glottis
cancer, a tumour confined to the glottis region with mobility of the vocal cords (ii) healthy patients and (iii) patients
under medical treatment, operated via endoscopic laser or traditional lancet technique. The sentence spoken by all of
them is the italian word \emph{aiuole} (flower-beds), since it contains a lot of vowels: Patients suffering from dysphonia
need a great effort in saying this word.

To classify the sentences we make use of the feature space introduced in \cite{fspace}, whose components are special 
quantities extracted from the time series: Spectral factor, pseudo-entropy, pseudo-correlation dimension, first zero-crossing 
of the autocorrelation function, first Lyapunov exponent, prediction error, jitter, shimmer, peaks in the phoneme transition, 
residual noise. It is not necessary here to go into details, one has just to know that this space represents the proper
object where a good classification of vocal pathologies is feasible.

\begin{figure}
\centerline{\psfig{file=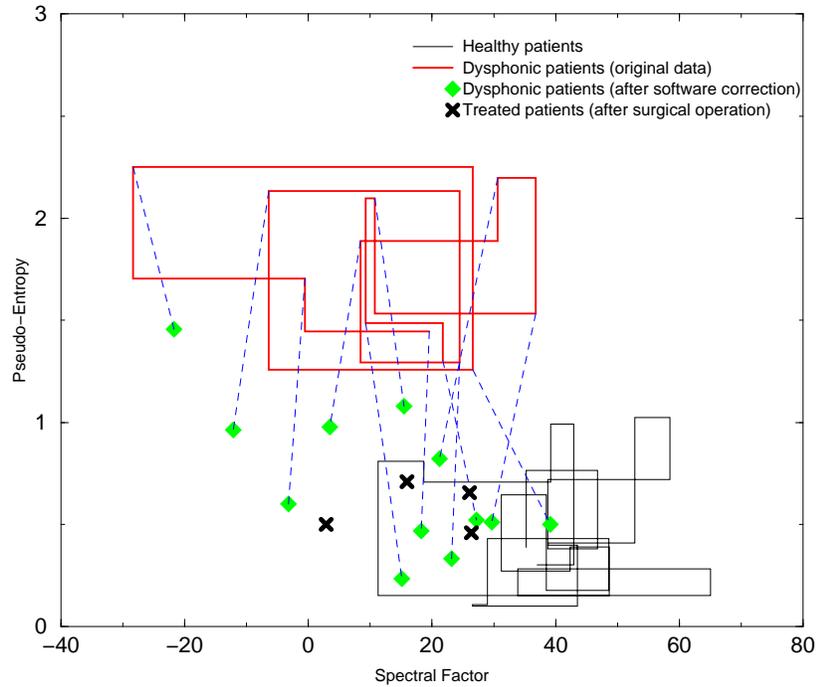,width=11cm,angle=270}}
\caption[]{\small\label{fig:corr4}
Two-dimensional projection of the feature space. Healthy patients are characterized by a pseudo-entropy close to 0.5 and a
spectral factor whitin the range [10,60]. Sick patients populate a region of the feature space where the pseudo-entropy
has a value close to 2 and the spectral factor ranges between -30 and 30. A bold cross refers to a patient after a
surgical operation. The effect of this treatment is a drastic decrease of the pseudo-entropy and a moderate increase in
the value of the spectral factor. Dotted lines link the points before the filtering to the same samples after the proposed
software corrections. The algorithm produces similar effects of a surgical treatment.}  
\end{figure}

In Fig.\ref{fig:corr4} we see a projection of the feature space onto the spectral factor and the pseudo-entropy dimensions; 
we can see how healthy and sick patients populate different regions and which are the results of the medical treatments 
performed on four patients (indicated by a bold cross).
Unfortunately we don't have samples from the same treated people before the surgical operation, 
so that it is not possible to link the four cases directly to the pre-operatory phase.
The dotted lines link the points before the filtering to the points after the attenuation of the noise-like features. 
We can force the algorithm to perform a stronger filtering of the dysphonic time series (as explained in the Appendix, this
is possible increasing the number of iteration of the correction routine), but one has always to be careful: 
In order to get a better value of some components of the feature space, one has to come to a compromise with some other 
quantities and the best answer comes, as usual in these cases, from a direct listening of the time series 
(after the convertion in a \emph{.wav} format).

The results shown in Fig.\ref{fig:corr4} have been obtained with the following set of parameters: $m = 30$, $\nu = 4$ and
$\epsilon = 0.3 \sigma$, with $\sigma$ being the variance of the original data. 
Dysphonic patients, in the (spectral factor, pseudo-entropy) plane, are spread around the average point (5,1.8); the centre
of mass of the healthy cluster is (35,0.5). The surgical treatment produces an average correction located in (20,0.6).
Our software moves the centre of mass of sick patients to (15,0.7).
In getting these results the value of $m$ and $\nu$ is not so
crucial, provided that $m\nu > l_p$, being $l_p$ the extension of a pitch. The program is more sensitive to the choice of
$\epsilon$, in the following way: The length of the dotted line of Fig.\ref{fig:corr4} is somehow proportional to the
value of $\epsilon$. Unfortunately the angle is not constant: This means that up to a certain value of $\epsilon$ all the
corrections act along the same direction, beyond that threshold they start to deteriorate the voice. In the extreme case
of $\epsilon$ as big as the full embedding space, the averaging procedure performed by Eq. \ref{eq:filtro} would destroy
almost completely the signal (producing a pseudo-entropy and a spectral factor close to zero).

\begin{figure}
\centerline{\psfig{file=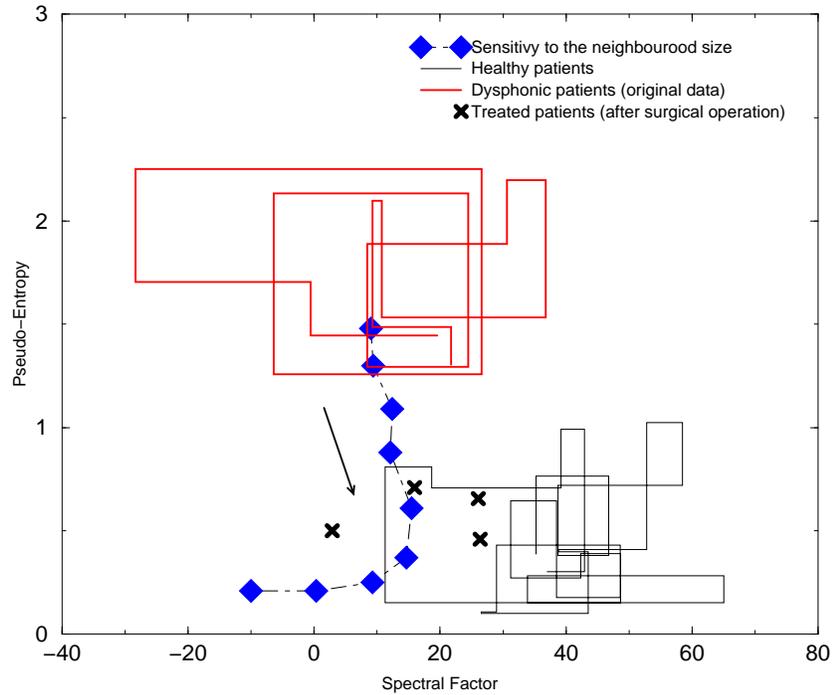,width=11cm,angle=270}}
\caption[]{\small\label{fig:correps}
Two-dimensional projection of the feature space. Healthy patients are characterized by a pseudo-entropy close to 0.5 and a
spectral factor whitin the range [10,60]. Sick patients populate a region of the feature space where the pseudo-entropy
has a value close to 2 and the spectral factor ranges between -30 and 30. The path followed by the bold diamands is the
effect of corrections with increasing $\epsilon$ ($\epsilon$ increases along the direction indicated by the arrow).
$\epsilon = 0$ coincides with the original sample, since no neighbour can be identified and therefore no correction is
performed. The last three points are the result of a filtering with a too big neighbourood size. Only for two values of
$\epsilon$ the corrected time series lies in the healthy region.}
\end{figure}

The sensitivity of the program to the choice of $\epsilon$ is illustrated in Fig. \ref{fig:correps}. There we have filtered
one sample with 8 different values of the neighbourood size. The original position is the one with the biggest value of the
pseudo-entropy. $\epsilon$ is increasing along the direction indicated by the arrow and only for two values of it the
corrected point lies in the healthy region. For the last three corrections the neighbourood size was absolutely too big.
The other samples behave in a qualitatively similar way.

\section*{Conclusions}

To summarize, we have proposed a procedure to correct numerically some kind of vocal disorders. The main ingredients
are the theory of dynamical systems, the delay reconstruction from a scalar measurement, the method of recurrence plot and
a nonlinear noise reduction scheme. We also make use of a feature space in order to visualize the results of the filtering 
procedure. 
A detailed description of how the idea has been implemented is also provided in Appendix. We have discussed the meaning,
implications, methodology and mathematical background of the most relevant parameters and the sensitivity of the program
to their choice.
The software corrections perform an improvement on the voice quality that is comparable to what a surgical 
operation is able to do. This suggests the idea of implementing the procedure in a physical device able to help people 
correcting their voice, without having to undergo a medical treatment.

\section*{Acknowledgements}

The authors would like to thank for fruitful collaborations Prof. Hanspeter Herzel about modeling of the voice, Priv. Doz. 
Holger Kantz, Priv. Doz. Thomas Schreiber and Dr. Rainer Hegger about Chaos and Noise Reduction, Prof. Cecilia Salimbeni and 
the Phoniatric Section of the Careggi Hospital in Florence about Vocal Disorders.

\vspace*{5cm}

\section*{Appendix: Implementation outline}

We present here the sketch of the subroutine \emph{\bf filter} written in a \emph{pseudo-C} code. We skip all the details
related to the syntax of the programming language for a better readability.

\begin{verbatim}

void show_options(char *progname)
{
  fprintf(stderr,"\t-m embedding dimension [Default: 5]\n");
  fprintf(stderr,"\t-T max distance in time [Default: no limit]\n");
  fprintf(stderr,"\t-d delay [Default: 1]\n");
  fprintf(stderr,"\t-r minimal neighbourhood size \n\t\t"
	  "[Default: (interval of data)/1000]\n");
  fprintf(stderr,"\t-R maximal neighborhoodsize[Default: not set]\n");
  fprintf(stderr,"\t-i # of iterations [Default: 1]\n");
}

\end{verbatim}

The input of the program has to be a scalar time series corresponding to the amplitude of the voice signal. Every conventional
ASCII file is accepted with the data in one column. The parameter $m$ is the embedding dimension (typical values in the range
$[20,40]$), $T$ is the portion of the time series where to look for neighbours: Since every pitch gets neighbours only within 
the same phoneme, the typical value of $T$ is 2000 (with a sampling rate of 22,05 kHz, 2000 points are almost 100 ms, 
approximately the length of a phoneme). The parameter $d$ is the time delay, usually in the range $[3,10]$ such that the
product $md$ is as big as the length of a pitch. The size of the neighbourood $\epsilon$ is specified through $r$ and $R$,
its minimal and maximal value respectively: $r$ should equal the noise level and $R$ is usually $1.5 r$.
With the parameter $i$, number of iterations, one can set the strength of the filtering. Tipically we use $i=1$.

\begin{verbatim}

unsigned long lfind_neighbors(long act,double eps)
{  
  k=(int)((searchdim-1)*delay);
  k1=(int)((searchdim/2)*delay);
  i=(int)(series[act-k]/eps)&ib;
  j=(int)(series[act-k1]/eps)&ib;
  n=(int)(series[act]/eps)&ib;
  
  for (i1=i-1;i1<=i+1;i1++) {
    i2=i1&ib;
    for (j1=j-1;j1<=j+1;j1++) {
      j2=j1&ib;
      for (n1=n-1;n1<=n+1;n1++) {
	element=box[i2][j2][n1&ib];
	while (element != -1) {
	  if (labs(act-element) < maxdist) {
	    dx=0.0;
	    for (k=0;k<searchdim;k++) {
	      k1= k*(int)delay;
	      dx += fabs(series[act-k1]-series[element-k1]);
	    }
	    if (dx/(double)searchdim <= eps) {
	      dist[nf]=dx;
	      flist[nf++]=element;
	    }
	  }
	  element=list[element];
	}
      }
    }
  }
  return nf;
}

\end{verbatim}

Two points in the embedding space are considered neighbours if \emph{(labs(act-element) $<$ maxdist)}. For every point we
build here a list containing all its neighbours. We do it firstly in a two dimensional space to speed up the process.
If two points are neighbours in a five dimensional space, this relation holds also in two dimension. Of course the
opposite is not true.

\begin{verbatim}

void correct(unsigned long n)
{  
  epsinv=1./eps;
  for (i=0;i<dim*delay;i++)
    hcor[i]=0.0;

  i=(int)(series[n-(dim-1)*delay]*epsinv)&ibox;
  j=(int)(series[n]*epsinv)&ibox;
  
  for (i1=i-1;i1<=i+1;i1++) {
    i2=i1&ibox;
    for (j1=j-1;j1<=j+1;j1++) {
      element=box[i2][j1&ibox];
      while (element != -1) {
	if (labs(n-element) < maxdist) {
	  for (k=0;k<dim;k++) {
	    k1=k*delay;
	    dx=fabs(series[n-k1]-series[element-k1]);
	    if (dx > eps)
	      break;
	  }
	  if (k == dim) {
	    flist[nfound++]=element;
	    for (k=0;k<dim*delay;k++)
	      hcor[k] += series[element-k];
	  }
	}
	element=list[element];
      }
    }
  }
  for (k=0;k<dim*delay;k++) {
    corr[n-k] += (hcor[k]=series[n-k]-hcor[k]/nfound);
    nf[n-k]++;
  }
  for (i=0;i<nfound;i++) {
    j=flist[i];
    for (k=0;k<dim*delay;k++) {
      trend[j-k] += hcor[k];
      tcount[j-k]++;
    }
  }
}

\end{verbatim}

The second to last \emph{for} iteration performs exactly what indicated in Eq.\ref{eq:filtro}. The variable \emph{nfound} is
$|U_{\epsilon}({\bf s}_{n_0})|$.

\begin{verbatim}

int main(int argc,char **argv)
{    
  series=(double*)get_series(infile,&length,exclude,column,1);

  rescale_data(series,length,&d_min,&d_max);
  
  resize_eps=0;
  for (iter=1;iter<=iterations;iter++) {
    epsilon=mineps;
    all_done=0;
    epscount=1;
    allfound=0;
    fprintf(stderr,"Starting iteration %d\n",iter);
    while(!all_done) {
      put_in_box(epsilon);
      all_done=1;
      for (n=(searchdim)*delay-1;n<length;n++)
	if (!ok[n]) {
	  nfound=lfind_neighbors(n,epsilon);
	  if (nfound >= minn) {
	    correct(n);
	    ok[n]=epscount;
	    if (epscount == 1)
	      resize_eps=1;
	    allfound++;
	  }
	  else
	    all_done=0;
	}
      fprintf(stderr,"Corrected %ld points with epsilon= %e\n",allfound,
	      epsilon*d_max);
      epsilon *= epsfac;
      epscount++;
      if (epsilon > maxeps)
	break;
    }
        
    sprintf(ofname,"%s.%d",outfile,iter);
    file=fopen(ofname,"w");
    fprintf(stderr,"Opened %s for writing\n\n",ofname);
    for (i=0;i<length;i++) {
      fprintf(file,"%e\n",series[i]*d_max+d_min);
      if (stdo && (iter == iterations))
	fprintf(stdout,"%e\n",series[i]*d_max+d_min);
    }
    fclose(file);
  }
  return 0;
}

\end{verbatim}

The flow is the following: Get the time series, rescale it to a proper interval, apply the filter (call to the function
\emph{correct}) a \emph{iter} number of times, print the corrected time series to a new data file.

\end{document}